\documentstyle[prd,preprint,aps]{revtex}
\begin{document}
\draft
\title{Limits on the
Dipole Moments
      of the $\tau$-Lepton via the Process $e^{+}e^{-}\rightarrow \tau^+ \tau^- \gamma$
      in a Left-Right Symmetric Model}

\author{A. Guti\'errez-Rodr\'{\i}guez $^{1}$, M. A. Hern\'andez-Ru\'{\i}z $^{2}$
       and  L.N. Luis-Noriega $^{1}$}
\address{(1) Facultad de F\'{\i}sica, Universidad Aut\'onoma de Zacatecas\\       
Apartado Postal C-580, 98068 Zacatecas, Zac. M\'exico.}

\address{(2) Facultad de Ciencias Qu\'{\i}micas, Universidad Aut\'onoma de Zacatecas\\
Apartado Postal 585, 98068 Zacatecas, Zac. M\'exico.}

\date{\today}
\maketitle
\begin{abstract}
Limits on the anomalous magnetic moment and the electric dipole
moment of the $\tau$ lepton are calculated through the reaction
$e^{+}e^{-}\rightarrow \tau^+ \tau^- \gamma$ at the $Z_1$-pole and
in the framework of a left-right symmetric model. The results are
based on the recent data reported by the L3 Collaboration at CERN LEP.
Due to the stringent limit of the model mixing angle $\phi$,
the effect of this angle on the dipole moments is quite small.

\end{abstract}

\narrowtext

\section{Introduction}

In the Standard Model (SM) \cite{S.L.Glashow}, the electromagnetic interactions
of each of the three charged leptons are identical. However, there is no
experimentally verified explanation for the existence of three generations of
leptons nor for why they have such different masses. New insight might be forthcoming
if the leptons were observed to have a substructure which could manifest itself
in deviations from the SM values for the anomalous magnetic or electric dipole
moments. The anomalous moments for the electron and muon have been measured with
very high precision \cite{Cohen} compared to those of tau for which there are only
upper limits \cite{Silverman,Escribano,Grifols,L3,OPAL}.

In general, a photon may couple to a tau through its electric charge, magnetic
dipole moment or electric dipole moment. This coupling may be parametrised
using a matrix element in which the usual $\gamma^\mu$ is replaced by a more
general Lorentz-invariant form

\[
\Gamma^\mu=F_1(q^2)\gamma^\mu+F_2(q^2)\frac{i}{2m_\tau}\sigma^{\mu\nu}q_\nu+F_3(q^2)\sigma^{\mu\nu}\gamma^5q_\nu,
\]

\noindent where $m_\tau$ is the mass of the $\tau$ lepton and $q=p'-p$ is the momentum
transfer. The $q^2$-dependent form-factors, $F_i(q^2)$, have familiar interpretations
for $q^2=0$: $F_1(0)\equiv Q_\tau$ is the electric charge; $F_2(0)\equiv a_\tau=(g-2)/2$
is the anomalous magnetic moment; and $F_3\equiv d_\tau /Q_\tau$, where $d_\tau$
is the electric dipole moment.

The analysis of radiative $\tau$ pair production provides a means to determine
the anomalous magnetic and electric dipole moments of the $\tau$ lepton at $q^2=0$.
An anomalous magnetic dipole moment at $q^2=0$ ($F_2(0)$) or an electric dipole
moment ($F_3(0)$) affects the total cross section for the process
$e^{+}e^{-}\rightarrow \tau^+ \tau^- \gamma$ as well as the shape of energy
and angular distributions of the three final state particles \cite{Grifols,Gau,Biebel}.
Previous experimental limits \cite{Grifols,L3,OPAL,Swain,Taylor} on $F_2(0)$ and
$F_3(0)$ have been based on approximate calculations of the
$e^{+}e^{-}\rightarrow \tau^+ \tau^- \gamma$ cross section and photon
energy distribution.

The first direct determination of the anomalous magnetic moment of the $\tau$
lepton, {\it i.e.} of $F_2(0)$, is due to Grifols and M\'endez using L3 data
\cite{Grifols}. They derived a limit for $F_2(q^2=0)\leq 0.11$ and
$F_{EDM}(q^2=0)\leq 6\times 10^{-16}e$cm at $q^2=0$. More recently, Escribano
and Mass\'o \cite{Escribano} have used electroweak data to find
$d_\tau \leq 1.1\times 10^{-17}e$cm and $-0.004\leq a_\tau \leq 0.006$
at the $2\sigma$ confidence level.

On the $Z_1$ peak, where a large number of $Z_1$ events are collected at
$e^{+}e^{-}$ colliders, one may hope to constrain or eventually measure the
anomalous magnetic moment and electric dipole moment of the $\tau$ by selecting
$\tau^{+}\tau^{-}$ events accompanied by a hard photon. The Feynman diagrams
which give the most important contribution to the cross section are shown in
Fig. 1.

Our aim in this paper is to analyze the reaction
$e^{+}e^{-}\rightarrow \tau^+ \tau^- \gamma$. We use recent data collected
with the L3 and OPAL detector at CERN LEP \cite{L3,OPAL} in the $Z_1$
boson resonance. The analysis is carried out in the context of a left-right
symmetric model \cite{R.N.Mohapatra,G.Senjanovic,G.Senjanovic1} and we attribute
a magnetic moment and an electric dipole moment to the tau lepton.
Processes measured in the resonance serve to set limits on the tau 
magnetic moment and electric dipole moment. We take advantage of this fact
to set limits for $a_\tau$ and $d_\tau$ for different values of the mixing
angle $\phi$ \cite{M.Maya,J.Polak,L3C}, which is consistent with other constraints
previously reported \cite{Escribano,Grifols,L3,OPAL}.

We do our analysis on the $Z_1$ peak ($s=M^{2}_{Z_1}$). Our results are
therefore independent of the mass of the additional heavy $Z_2$ gauge
boson which appears in these kind of models. Thus, we have the mixing
angle $\phi$ between the left and the right bosons as the only additional
parameter apart from the SM parameters.

This paper is organized as follows: In Sect. II we describe the model with
the Higgs sector having two doublets and one bidoublet. In Sect. III we 
present the calculus of the process
$e^{+}e^{-}\rightarrow \tau^+ \tau^- \gamma$. In Sect. IV we make the
numerical computations. Finally, we summarize our results in Sect. V.


\section{The Left-Right Symmetric Model (LRSM)}

We consider a Left-Right Symmetric Model (LRSM) consisting of one bidoublet
$\Phi$ and two doublets $\chi _{L}$, $\chi _{R}$. The vacuum expectation
values of $\chi _{L}$, $\chi _{R}$ break the gauge symmetry to give mass to the left and right heavy
gauge bosons. This is the origin of the parity violation at low energies
\cite{R.N.Mohapatra} {\it i.e.}, at energies produced in actual accelerators.
The Lagrangian for the Higgs sector of the LRSM is \cite{G.Senjanovic}

\begin{equation}
{\cal L}_{LRSM}=(D_{\mu}\chi _{L})^{\dagger}(D^{\mu}\chi _{L})
+ (D_{\mu}\chi _{R})^{\dagger}(D^{\mu}\chi _{R}) + Tr(D_{\mu}\Phi)^{\dagger}(D^{\mu}\Phi).
\end{equation}

\noindent The covariant derivatives are written as

\begin{eqnarray}
D_{\mu}\chi _{L}&=&\partial _{\mu}\chi _{L}-\frac{1}{2}ig{\bf \tau}\cdot {\bf W}_{L}\chi _{L}
-\frac{1}{2}ig^{'}B\chi _{L},\nonumber\\
D_{\mu}\chi _{R}&=&\partial _{\mu}\chi _{R}-\frac{1}{2}ig{\bf \tau}\cdot {\bf W}_{R}\chi _{R}
-\frac{1}{2}ig^{'}B\chi _{R},\\
D_{\mu}\Phi&=&\partial _{\mu}\Phi-\frac{1}{2}ig({\bf \tau}\cdot {\bf W}_{L}\Phi
-\Phi {\bf \tau}\cdot {\bf W}_{R}).\nonumber
\end{eqnarray}

There are seven gauge bosons in this model: the charged $W^{1}_{L,R}$, $W^{2}_{L,R}$
and the neutral $W^{3}_{L,R}$, $B$. The gauge couplings constants $g_L$ and $g_R$
of the $SU(2)_L$ and $SU(2)_R$ subgroups, respectively, are equal: $g_{L} = g_{R}=g$,
since manifest left-right symmetry is assumed \cite{M.A.B}. $g'$ is the gauge
coupling for the $U(1)$ group.
 
The transformation properties of the Higgs bosons under the group $SU(2)_{L}\times SU(2)_{R}\times U(1)$
are $\chi _{L} \sim (1/2, 0, 1)$, $\chi _{R} \sim (0, 1/2, 1)$ and $\Phi \sim (1/2, 1/2^{*}, 0)$.
After spontaneous symmetry breaking, the ground states are of the form

\begin{equation}
\langle \chi _{L} \rangle = \frac{1}{\sqrt{2}}\left( 
					      \begin{array}{c}
					      0\\ 
					      v_{L}
					      \end{array}
					      \right), \hspace*{2mm}
\langle \chi_{R} \rangle = \frac{1}{\sqrt{2}}\left(                           
					     \begin{array}{c} 
					      0\\ 
					      v_{R}
					      \end{array}
					      \right), \hspace*{2mm}
\langle \Phi \rangle = \frac{1}{\sqrt{2}}\left(
					 \begin{array}{ll}
					 k&0\\
                                         0&k'
					 \end{array}
					 \right),
\end{equation}

\noindent which break the symmetry group to form the $U(1)_{em}$, giving mass
to the gauge bosons and fermions with the photon remaining massless.
In Eq. (3), $v_L$, $v_R$, $k$ and $k'$ are the vacuum expectation values.
The part of the Lagrangian that contains the mass terms for the charged
boson is

\begin{equation}
{\cal L}^{C}_{mass} = (W^{+}_{L}\hspace*{2mm} W^{+}_{R}) M^{C}\left(
						\begin{array}{c}
						W^{-}_{L}\\
						W^{-}_{R}
						\end{array}
						\right),
\end{equation}

\noindent where $W^{\pm} = \frac{1}{\sqrt{2}}(W^{1} \mp W^{2})$.

The mass matrix $M^{C}$ is

\begin{equation}
M^{C} = \frac{g^{2}}{4}\left(
		       \begin{array}{cc}
		       v^{2}_{L}+k^{2}+k^{'2}&-2kk^{'}\\
		       -2kk^{'}&v^{2}_{R}+k^{2}+k^{'2}
		       \end{array}
                       \right).
\end{equation}

\noindent This matrix is diagonalized by an orthogonal transformation 
parametrized \cite{M.A.B} by the angle $\zeta$. This angle has a 
very small value because of the hyperon $\beta$ decay data \cite{M.Aquino}.

Similarly, the part of the Lagrangian that contains the mass terms for the neutral
bosons is

\begin{equation}
{\cal L}^{N}_{mass} = \frac{1}{8}(W^{3}_{L} \hspace*{2mm}  W^{3}_{R} \hspace*{2mm} B)
		      M^{N}\left(
			   \begin{array}{c}
			   W^{3}_{L}\\
			   W^{3}_{R}\\
			   B
			   \end{array}
			   \right),
\end{equation}

\noindent where the matrix $M^{N}$ is given by

\begin{equation}
M^{N} = \frac{1}{4}\left(
		   \begin{array}{ccc}
		   g^{2}(v^{2}_{L}+k^{2}+k^{'2})&-g^{2}(k^{2}+k^{'2})&-gg^{'}v^{2}_{L}\\
		   -g^{2}(k^{2}+k^{'2})&g^{2}(v^{2}_{R}+k^{2}+k^{'2})&-gg^{'}v^{2}_{R}\\
		   -gg^{'}v^{2}_{L}&-gg^{'}v^{2}_{R}&g^{'2}(v^{2}_{L}+v^{2}_{R})
		   \end{array}
		   \right).
\end{equation}

Since the process $e^{+}e^{-}\rightarrow \nu \bar \nu \gamma$ is neutral, we center our 
attention on the mass terms of the Lagrangian for the neutral sector as shown
in Eq. (6).

The matrix $M^{N}$ for the neutral gauge bosons is diagonalized by an orthogonal
transformation which can be written in terms of the angles $\theta _{W}$ and $\phi$ \cite{J.Polak1}

\begin{equation}
U^{N} = \left(
	\begin{array}{ccc}
	c_{W}c_{\phi}&-s_{W}t_{W}c_{\phi}-r_{W}s_{\phi}/c_{W}&t_{W}(s_{\phi}-r_{W}c_{\phi})\\
	c_{W}s_{\phi}&-s_{W}t_{W}s_{\phi}+r_{W}c_{\phi}/c_{W}&-t_{W}(c_{\phi}+r_{W}s_{\phi})\\
	s_{W}&s_{W}&r_{W}
	\end{array}
	\right),
\end{equation}

\noindent where $c_{W}=\cos \theta _{W}$, $s_{W}=\sin \theta _{W}$, $t_{W}=\tan\theta_{W}$
and $r_{W}=\sqrt{\cos 2\theta_{W}}$, and $\theta _{W}$ is the electroweak mixing angle.
Here, $c_{\phi}=\cos\phi$ and $s_{\phi}=\sin\phi$. The angle $\phi$ is considered
as the angle that mixes the left and right handed neutral gauge bosons $W^{3}_{L,R}$.
The expression that relates the left and right handed neutral gauge bosons $W^{3}_{L,R}$
and $B$ with the physical bosons $Z_{1}$, $Z_{2}$ and the photon is:

\begin{equation}
\left(
\begin{array}{c}
Z_{1}\\
Z_{2}\\
A
\end{array}
\right)
=
U^{N}\left(
     \begin{array}{c}
     W^{3}_{L}\\
     W^{3}_{R}\\
     B
     \end{array}
     \right).
\end{equation}

The diagonalization of (5) and (7) gives the mass of the charged
$W^{\pm}_{1,2}$ and neutral $Z_{1,2}$ physical fields:

\begin{equation}
M^{2}_{W_{1,2}}=\frac{g^{2}}{8}[v^{2}_{L}+v^{2}_{R}+2(k^{2}+k^{'2})
\mp \sqrt{(v^{2}_{R}-v^{2}_{L})^{2}+16(kk^{'})^{2}}],
\end{equation}

\begin{equation}
M^{2}_{Z_{1},Z_{2}}=B\mp \sqrt{B^{2}-4C},
\end{equation}

\noindent respectively, with

\[
B=\frac{1}{8}[(g^{2}+g^{'2})(v^{2}_{L}+v^{2}_{R})+2g^{2}(k^{2}+k^{'2})],
\]

\[
C=\frac{1}{64}g^{2}(g^{2}+2g^{'2})[v^{2}_{L}v^{2}_{R}+(k^{2}+k^{'2})(v^{2}_{L}+v^{2}_{R})].
\]

Taking into account  that $M^{2}_{W_{2}}\gg M^{2}_{W_{1}}$, we conclude from
the expressions for the masses of $M_{Z_{1}}$ and $M_{Z_{2}}$, that the relation
$M^{2}_{W_{1}}=M^{2}_{Z_{1}}\cos ^{2}\theta_{W}$ still holds in this model.

From the Lagrangian of the LRSM, we extract the terms for the neutral interaction
of a fermion with the gauge bosons $W^{3}_{L,R}$ and $B$:

\begin{equation}
{\cal L}^{N}_{int}=g(J^{3}_{L}W^{3}_{L}+J^{3}_{R}W^{3}_{R})+\frac{g^{'}}{2}J_{Y}B.
\end{equation}

Specifically, the Lagrangian interaction for $Z_{1}\rightarrow f\bar f$ \cite{P.Langacker} is

\begin{equation}
{\cal L}^{N}_{int}=\frac{g}{c_{W}}Z_{1}[(c_{\phi}-\frac{s^{2}_{W}}{r_{W}}s_{\phi})J_{L}    
		   -\frac{c^{2}_{W}}{r_{W}}s_{\phi}J_{R}],
\end{equation}

\noindent where the left (right) current for the fermions are

\[
J_{L,R}=J^{3}_{L,R}-\sin ^{2}\theta_{W}J_{em}
\]

\noindent and

\[
J_{em}=J^{3}_{L}+J^{3}_{R}+\frac{1}{2}J_{Y}
\]

\noindent is the electromagnetic current. From (13), we find that the amplitude
${\cal M}$ for the decay of the $Z_{1}$ boson with polarization $\epsilon ^{\lambda}$
into a fermion-antifermion pair is:

\begin{equation}
{\cal M}=\frac{g}{c_{W}}[\bar u \gamma^{\mu}\frac{1}{2}(ag_{V}-bg_{A}\gamma_{5})v]\epsilon^{\lambda}_{\mu},
\end{equation}

\noindent with

\begin{equation}
a=c_{\phi}-\frac{s_{\phi}}{r_{W}} \hspace{5mm} \mbox{and} \hspace{5mm} b=c_{\phi}+r_{W}s_{\phi},
\end{equation}

\noindent where $\phi$ is the mixing parameter of the LRSM \cite{M.Maya,J.Polak}.

In the following section, we make the calculations for the reaction
$e^{+}e^{-}\rightarrow \tau^+ \tau^- \gamma$ by using the expression (14)
for the transition amplitude.

\section{The Total Cross Section}

We calculate the total cross section of the process $e^{+}e^{-}\rightarrow \tau^+ \tau^- \gamma$
using the Breit-Wigner resonance form \cite{Data2002,Renton}:

\begin{equation}
\sigma(e^+e^-\rightarrow \tau^+ \tau^- \gamma)=\frac{4\pi(2J+1)\Gamma_{e^+e^-}\Gamma_{\tau^+ \tau^- \gamma}}{(s-M^2_{Z_{1}})^2+M^2_{Z_{1}}\Gamma^2_{Z_{1}}},
\end{equation}

\noindent where $\Gamma_{e^+e^-}$ is the decay rate of $Z_1$ to the channel
$Z_1 \to e^+e^-$ and $\Gamma_{\tau^+ \tau^- \gamma}$ is the decay rate of $Z_1$
to the channel $Z_1 \rightarrow \tau^+ \tau^- \gamma$.

In the next subsection, we calculate the widths of Eq. (16).

\subsection{Width of $Z_1 \to e^+e^-$}

We now calculate the total width of the reaction

\begin{equation}
Z_1 \to e^+e^-,
\end{equation}

\noindent in the context of the left-right symmetric model which is described
in Section II.

The expression for the total width of the process $Z_1 \to e^+e^-$,
according to the diagrams depicted in Fig. 1 and using the expression
for the amplitude given in Eq. (14), is:

\begin{equation}
\Gamma_{(Z_1 \to e^+e^-)}=\frac{G_FM^3_{Z_{1}}}{6\pi \sqrt{2}}\sqrt{1-4\eta}
[a^2(g^e_V)^2(1+2\eta)+b^2(g^e_A)^2(1-4\eta)],
\end{equation}

\noindent where $\eta=m^2_e/M^2_{Z_{1}}$.

\noindent We take $g^e_V=-\frac{1}{2}+2\sin^2\theta_W$ and $g^e_A=-\frac{1}{2}$,
from the experimental data \cite{Data2002} and $m_e=0$ so that the total
width is

\begin{equation}
\Gamma_{(Z_1 \to e^+e^-)}=\frac{\alpha M_{Z_{1}}}{24}
[\frac{\frac{1}{2}(a^2+b^2)-4a^2x_W+8a^2x^2_W}{ x_W(1-x_W)}],
\end{equation}

\noindent where $x_W=\sin^2\theta_W$ and $\alpha =e^2/4\pi$ is the fine structure
constant.

\subsection{Width of $Z_1 \rightarrow \tau^+ \tau^- \gamma$}

The expression for the Feynman amplitude ${\cal M}$ of the process $Z_1 \rightarrow  \tau^+\tau^- \gamma$
is due only to the $Z_1$ boson exchange, as shown in the diagrams in
Fig. 1. We use the expression for the amplitude given in Eq. (14) and assume that a
tau lepton is characterized by the following phenomenological parameters:
a charge radius $\langle r^{2}\rangle$, a magnetic moment $a_\tau$,
and an electric dipole moment $d_\tau$. Therefore, the expression for the Feynman
amplitude ${\cal M}$ of the process $Z_1 \rightarrow  \tau^+\tau^- \gamma$ is
given by

\begin{eqnarray}
{\cal M}_{1}&=&[\bar u(p_{\tau^-})\Gamma^{\alpha}\frac{i}{(\ell\llap{/}-m_{\tau})}
(-\frac{ig}{2\cos\theta_W}\gamma^{\beta}(a g^\tau_V-bg^\tau_A\gamma_{5}))v(p_{\tau^+})]
\epsilon^\lambda_\alpha(\gamma) \epsilon^\lambda_\beta(Z_1)
\end{eqnarray}

\noindent and

\begin{eqnarray}
{\cal M}_{2}&=&[\bar u(p_{\tau^-})(-\frac{ig}{2\cos\theta_W}\gamma^{\beta}(ag^\tau_V-bg^\tau_A\gamma_{5}))
\frac{i}{(k\llap{/}-m_{\tau})}\Gamma^{\alpha}v(p_{\tau^+})]
\epsilon^\lambda_\alpha(\gamma) \epsilon^\lambda_\beta(Z_1),
\end{eqnarray}

\noindent where

\begin{equation}
\Gamma^{\alpha}=eF_{1}(q^{2})\gamma^{\alpha}+\frac{ie}{2m_{\tau}}F_{2}(q^{2})\sigma^{\alpha \mu}q_{\mu}+ eF_{3}(q^{2})\gamma_5\sigma^{\alpha \mu}q_{\mu}           
\end{equation}

\noindent is the tau electromagnetic vertex, $e$ is the charge of the electron,
$q^\mu$ is the photon momentum and $F_{1,2,3}(q^2)$ are the electromagnetic
form factors of the tau which correspond to charge radius, magnetic moment
and electric dipole moment, respectively, at $q^2=0$ \cite{Escribano,Grifols}.
$\epsilon^{\lambda}_{\alpha}$ and  $\epsilon^{\lambda}_{\beta}$
are the polarization vectors of photon and of the boson $Z_1$, respectively.
$l$ ($k$) stands by the momentum of the virtual tau (antitau),
and the coupling constants $a$ and $b$ are given in the Eq. (15).

After applying some of the trace theorems of the Dirac matrices and of
sum and average over the initial and final spin, the square of the matrix
elements becomes

\begin{equation}
\sum_s\mid{\cal M}_T\mid^2=\frac{g^2}{\cos^2\theta_W}[\frac{e^2a^2_\tau}{4m^2_\tau}+d^2_\tau]
[(a^2(g^\tau_V)^2+b^2(g^\tau_A)^2)(s-2\sqrt{s}E_\gamma)
+b^2(g^\tau_A)^2E^2_\gamma\sin^2\theta_\gamma].
\end{equation}

Now that we know the square of the Eq. (23) transition amplitude, our
final step is to calculate the total width of $Z_1 \rightarrow \tau^+ \tau^- \gamma$:

\begin{eqnarray}
\Gamma_{(Z_1 \rightarrow \tau^+ \tau^- \gamma)}&=&\int\frac{\alpha}{12\pi^2M_{Z_{1}}x_W(1-x_W)}[\frac{e^2a^2_\tau}{4m^2_\tau}+ d^2_\tau]\nonumber\\
&&[(a^2(g^\tau_V)^2+b^2(g^\tau_A)^2)(s-2\sqrt{s}E_\gamma)+b^2(g^\tau_A)^2E^2_\gamma\sin^2\theta_\gamma]E_\gamma dE_\gamma d\cos\theta_\gamma,
\end{eqnarray}

\noindent where $E_{\gamma}$ and $\cos\theta_{\gamma}$ are the energy and
scattering angle of the photon.

The substitution of (19) and (24) in (16) gives

\begin{eqnarray}
\sigma(e^{+}e^{-}\rightarrow \tau^+ \tau^-\gamma)&=&\int\frac{\alpha^2}{48\pi}
[\frac{e^2a^2_\tau}{4m^2_\tau}+d^2_\tau]
[\frac{\frac{1}{2}(a^2+b^2)-4a^2x_W+8a^2x^2_W}{x^2_W(1-x_W)^2}]\nonumber\\
&&[\frac{[\frac{1}{2}(a^2+b^2)-4a^2x_W+8a^2x^2_W]
(s-2\sqrt{s}E_\gamma)+\frac{1}{2}b^2E^2_\gamma\sin^2\theta_\gamma}{(s-M^2_{Z_{1}})^2+M^2_{Z_{1}}\Gamma^2_{Z_{1}}}]\nonumber\\
&&E_\gamma dE_\gamma d\cos\theta_\gamma,
\end{eqnarray}

\noindent where $x_W\equiv \sin^2\theta_W$.

After evaluating the limit when the mixing angle is $\phi=0$, the expression for
$a$ and $b$ is reduced to $a=b=1$ and Eq. (25) is reduced to the expression (4)
given in Ref. \cite{Grifols}.

\section{Results}

In practice, detector geometry imposes a cut on the photon polar angle with
respect to the electron direction, and further cuts must be applied on the
photon energy and minimum opening angle between the photon and tau in order
to suppress background from tau decay products. Therefore, to evaluate the
integral of the total cross section as a function of mixing angle $\phi$,
we require cuts on the photon angle and energy to avoid divergences when the
integral is evaluated at the important intervals of each experiment. We
integrate over $\cos\theta_\gamma$ from $-0.74$ to $0.74$ and
$E_\gamma$ from 5 $GeV$ to 45.5 $GeV$ for various fixed values
of the mixing angle $\phi =-0.009, -0.005, 0, 0.004$. Using the 
numerical values: $\sin^2\theta_W=0.2314$, $m_\tau=1.776$ $GeV$,
$M_{Z_1}=91.187$ $GeV$, and $\Gamma_{Z_1}=2.49$ $GeV$, we obtain the cross section
$\sigma=\sigma(\phi, a_\tau,d_\tau)$.

For the mixing angle $\phi$ between $Z_{1}$ and $Z_{2}$, we use the reported
data of M. Maya {\it et al.} [17]:

\begin{equation}
-9\times 10^{-3}\leq \phi \leq 4\times 10^{-3},
\end{equation}

\noindent with a $90$ $\%$ C. L. Other limits on the mixing angle $\phi$ reported
in the literature are given in the Refs. [18,19].

As was discussed in Refs. \cite{L3,OPAL}, $N\approx\sigma(\phi, a_\tau, d_\tau){\cal L}$.
Using the Poisson statistic \cite{L3,Barnett}, we require that $N\approx\sigma(\phi, a_\tau, d_\tau){\cal L}$
be less than 1559, with ${\cal L}= 100$ $pb^{-1}$, according to the data
reported by the L3 Collaboration Ref. \cite{L3}. Taking this into consideration,
we put a bound for the tau lepton magnetic moment as a function of the $\phi$
mixing parameter with $d_\tau=0$. We show the value of this bound for values
of the $\phi$ parameter in Tables 1 and 2.\\

\begin{center}
\begin{tabular}{|c|c|c|}\hline
$\phi$&$a_\tau$&$d_\tau (10^{-16}e \mbox{cm})$\\ \hline
\hline
-0.009&0.084&4.61\\
\hline
-0.005&0.083&4.59\\
\hline
0&0.082&4.55\\
\hline
0.004&0.081&4.53\\
\hline
\end{tabular}
\end{center}

\begin{center}
Table 1. Limits on the $a_\tau$ magnetic moment and $d_\tau$
electric dipole moment of the $\tau$-lepton for different values of the
mixing angle $\phi$ before the $Z_1$ resonance, {\it i.e.}, $s \approx M^2_{Z_1}$ .
\end{center}

These results differ from the limits obtained in the references
\cite{L3,OPAL}. However, the derived limits in Table 1 could be improved by
including data from the entire $Z_1$ resonance as shown in Table 2.\\

\begin{center}
\begin{tabular}{|c|c|c|}\hline
$\phi$&$a_\tau$&$d_\tau(10^{-16}e \mbox{cm})$\\ \hline
\hline
-0.009&0.06&3.27\\
\hline
-0.005&0.059&3.25\\
\hline
0&0.058&3.22\\
\hline
0.004&0.057&3.21\\
\hline
\end{tabular}
\end{center}

\begin{center}
Table 2. Limits on the $a_\tau$ magnetic moment and $d_\tau$
electric dipole moment of the $\tau$-lepton for different values of the
mixing angle $\phi$ in the $Z_1$ resonance, {\it i.e.}, $s=M^2_{Z_1}$ .
\end{center}

The above analysis and comments can readily be translated to the electric
dipole moment of the $\tau$-lepton. The resulting limit for the electric
dipole moment as a function of the $\phi$ mixing parameter is shown in Table 1.

The results in Table 2 for the electric dipole moment are in agreement
with those found by the L3 Collaboration \cite{L3}.

Fig. 2 shows the total cross section as a function of the mixing angle
$\phi$ for the limits of the magnetic moment given in Tables 1, 2. We observe
in Fig. 2 that for $\phi=0$, we reproduce the data previously reported in the
literature. Also, we observe that the total cross section increases constantly
and reaches its maximum value for $\phi = 0.004$.

We end this section by plotting the differential cross section in Fig. 3 as
a function of the photon energy versus the cosine of the opening angle between
the photon and the beam direction ($\theta_\gamma$), for $\phi = 0.004$ and
$a_\tau = 0.057$. We observe in this figure that the energy and angular
distributions are consistent with those reported in the literature. In addition,
the form of the distributions does not change significanty for the values 
$\phi$ and $a_\tau$ because $\phi$ and $a_\tau$ are very small in value,
as shown in Table 2.

\section{Conclusions}

We have determined a limit on the magnetic moment and the electric dipole moment
of the tau lepton in the framework of a left-right symmetric model as
a function of the mixing angle $\phi$, as shown in Table 1 and Table 2.

In summary, we conclude that the estimated limit for the tau lepton magnetic
moment and the electric dipole moment are almost independent of the experimental
allowed values of the $\phi$ parameter of the model. In the limit $\phi = 0$,
our bound takes the value previously reported in the literature \cite{L3,OPAL}.\\


\begin{center}
{\bf Acknowledgments}
\end{center}

This work was supported in part by {\bf SEP-CONACYT} (M\'exico)
({\bf Projects: 2003-01-32-001-057, 40729-F}), {\it Sistema Nacional de Investigadores}
({\bf SNI}) (M\'exico) and {\it Programa de Mejoramiento al Profesorado} ({\bf PROMEP}).

\newpage

\begin{center}
FIGURE CAPTIONS
\end{center}

\bigskip

\noindent {\bf Fig. 1} The Feynman diagrams contributing to the process $e^{+}e^{-}\rightarrow \tau^+ \tau^- \gamma$
in a left-right symmetric model.

\bigskip

\noindent {\bf Fig. 2} The total cross section for $e^{+}e^{-}\rightarrow \tau^+ \tau^- \gamma$ 
as a function of $\phi$ and $a_\tau$ (Tables 1, 2).

\bigskip

\noindent {\bf Fig. 3} The differential cross section for $e^{+}e^{-}\rightarrow \tau^+ \tau^- \gamma$
as a function of $E_\gamma$ and $cos\theta_\gamma$ for $\phi=0.004$ and $a_\tau=0.057$.


\begin{references}

\bibitem{S.L.Glashow} S. L. Glashow, Nucl. Phys. {\bf 22}, (1961) 579; S. Weinberg, Phys. Rev.
                      Lett. {\bf 19}, (1967) 1264; A. Salam, in Elementary Particle Theory,
                      Ed. N. Svartholm (Almquist and Wiskell, Stockholm, 1968) 367.

\bibitem{Cohen} E.R. Cohen and B.N. Taylor, Rev. Mod. Phys. {\bf 59}, (1987) 1121.

\bibitem{Silverman} D.J. Silverman and G.L. Shaw, Phys. Rev. {\bf D27}, (1983) 1196.

\bibitem{Escribano} R. Escribano and E. Mass\'o, Phys. Lett. {\bf B395}, (1997) 369.

\bibitem{Grifols} J.A. Grifols and A. M\'endez, Phys. Lett. {\bf B255}, (1991) 611;
                  Erratum {\it ibid} {\bf B259}, (1991) 512.

\bibitem{L3} The L3 Collaboration, Phys. Lett. {\bf B434}, (1998) 169, and references therein.

\bibitem{OPAL} The OPAL Collaboration, Phys. Lett. {\bf B431}, (1998) 188, and references therein.  

\bibitem{Gau} S.S. Gau, T. Paul, J. Swain, L. Taylor, Nucl. Plys. {\bf B523}, (1998) 439.

\bibitem{Biebel} J. Biebel and T Riemann, Z. Phys. {\bf C76}, (1997) 53.

\bibitem{Swain} J. Swain, Nucl. Phys. {\bf B} (Proc. Suppl.) {\bf 98}, (2001) 351.

\bibitem{Taylor} L. Taylor, Nucl. Phys. {\bf B} (Proc. Suppl.) {\bf 76}, (1999) 237.

\bibitem{R.N.Mohapatra} R. N. Mohapatra, Prog. Part. Nucl. Phys. {\bf 26}, (1992) 1.

\bibitem{G.Senjanovic} G. Senjanovic, Nucl. Phys. {\bf B 153}, (1979) 334.

\bibitem{G.Senjanovic1} G. Senjanovic and R. N. Mohapatra, Phys. Rev. {\bf D12}, (1975) 1502.

\bibitem{M.Maya} M. Maya and O. G. Miranda, Z. Phys. {\bf C68}, (1995) 481.

\bibitem{J.Polak} J. Polak, M. Zralek, Phys. Rev. {\bf D46}, (1992) 3871.

\bibitem{L3C} L3 Collab., O. Adriani {\it et al.}, Phys. Lett. {\bf B306}, (1993) 187.

\bibitem{M.A.B} M. A. B. Beg, R. V. Budny, R. Mohapatra, A. Sirlin, Phys. Rev. Lett. {\bf 22}, (1977) 1252.

\bibitem{M.Aquino} M. Aquino, A. Fern\'andez, A. Garc\'{\i}a, Phys. Lett. {\bf B261}, (1991) 280.

\bibitem{J.Polak1}J. Polak, M. Zralek, Nucl. Phys. {\bf B363}, (1991) 385.

\bibitem{P.Langacker} P. Langacker, M. Luo, A. K. Mann, Rev. Mod. Phys. {\bf 64}, (1992) 87.

\bibitem{Data2002} Particle Data Group, Phys. Rev. {\bf D66}, (2002) 1.

\bibitem{Renton} Peter Renton, "{\it An Introduction to the Physics of Quarks and Leptons}",
                  Cambridge University Press, 1990.

\bibitem{Barnett} R. M. Barnett {\it et al.} Phys. Rev. {\bf D54}, (1996) 166.

\end{references}
\end{document}